\documentclass[aps,prl,twocolumn,superscriptaddress]{revtex4}

\usepackage{amsmath}
\usepackage{amsfonts}
\usepackage{amssymb}
\usepackage{epsfig}
\usepackage{hyperref}
\usepackage{graphicx}
\usepackage{wrapfig}
\usepackage{bm}
\usepackage{upgreek} 
\usepackage{cancel} 
\usepackage[caption=false]{subfig}
\usepackage{float}
\usepackage{diagbox}

\begin{document}

\title{Heisenberg-like and Fisher-information-based uncertainty relations for $N$-electron $d$-dimensional systems}

\author{I.V. Toranzo}
\affiliation{Departamento de F\'{\i}sica At\'{o}mica, Molecular y Nuclear, Universidad de Granada, Granada 18071, Spain}
\affiliation{Instituto Carlos I de F\'{\i}sica Te\'orica y Computacional, Universidad de Granada, Granada 18071, Spain}

\author{S. L\'opez-Rosa}
\email[]{slopezrosa@us.es}
\affiliation{Instituto Carlos I de F\'{\i}sica Te\'orica y Computacional, Universidad de Granada, Granada 18071, Spain}
\affiliation{Departamento de F\'{\i}sica Aplicada II, Universidad de Sevilla, Sevilla 41012, Spain}

\author{R. O. Esquivel}
\email[]{esquivel@xanum.uam.mx}
\affiliation{Instituto Carlos I de F\'{\i}sica Te\'orica y Computacional, Universidad de Granada, Granada 18071, Spain}
\affiliation{Departamento de Qu\'imica, Universidad Aut\'onoma Metropolitana, M\'exico 09340, M\'exico}

\author{J.S. Dehesa}
\email[]{dehesa@ugr.es}
\affiliation{Departamento de F\'{\i}sica At\'{o}mica, Molecular y Nuclear, Universidad de Granada, Granada 18071, Spain}
\affiliation{Instituto Carlos I de F\'{\i}sica Te\'orica y Computacional, Universidad de Granada, Granada 18071, Spain}
\date{\today}

\begin{abstract}
Heisenberg-like and Fisher-information-based uncertainty relations which extend and generalize previous similar expressions are obtained for $N$-fermion $d$-dimensional systems. The contributions of both spatial and spin degrees of freedom are taken into account. The accuracy of some of these generalized spinned uncertainty-like relations is numerically examined for a large number of atomic and molecular systems.
\end{abstract}



\maketitle

\section{Introduction}

According to the density functional theory, the physical and chemical properties of atoms and molecules can be described in principle by means of functionals of the position electron density $\rho(\vec{r})$ and/or functionals of the momentum electron density  $\gamma(\vec{p})$ \cite{Parr-Yang, Sukumar, Lehtola}. Moreover, the qualitative and quantitative understanding of the electronic structure of atoms and molecules require in practice the knowledge of the expressions of the position and momentum space representations of the relevant physico-chemical quantities of these systems \cite{Thakkar1, Thakkar2}. These quantities can be fully determined by the position ordinary and frequency or entropic moments which for $d$-dimensional systems are given by
\begin{eqnarray}
\label{eq:avevalerad}
\langle r^{k}\rangle &=& \int_{\mathbb{R}_{d}} r^{k}\rho(\vec{r})\, d^{d}r,\\
\label{eq:avevalmom}
 W_{q}[\rho] &=& \int_{\mathbb{R}_{d}} \rho^{q}(\vec{r})\, d^{d}r
\end{eqnarray}
respectively, under certain conditions. A similar statement can be said for the momentum density $\gamma(\vec{p})$ in terms of the corresponding momentum moments $\langle p^{k}\rangle $ and $ Z_{q}[\gamma]$. The notation $r = |\vec{r}|$ and $p = |\vec{p}|$ is used throughout the paper.

The connections between these moments in the two conjugate position and momentum spaces are very important for both fundamental and practical reasons. Indeed, the position-momentum uncertainty principle for quantum systems that generalizes the seminal variance-based formulation of Heisenberg can be expressed in a more accurate and useful manner by use of ordinary moments of order higher than 2 \cite{gadre1, Angulo1, Zozor, Porras1, Porras2} and/or by means of entropic moments \cite{maassen}. On the other hand, numerous physical and chemical properties can be expressed in terms of some ordinary and entropic moments in both position and momentum representations \cite{epstein73, gadre1, Thakkar1}. Indeed, they describe and/or are closely related to some fundamental and/or experimentally accesible quantities, such as the diamagnetic susceptibility ($\langle r^{2}\rangle$), the kinetic energy ($\langle p^{2}\rangle$), the Thomas-Fermi kinetic energy ($W_{5/3}$), the Dirac-Slater exchange energy ($W_{4/3}$, $\langle p\rangle$), the height peak of the Compton profile ($\langle p^{-1}\rangle$), the relativistic Breit-Pauli energy ($\langle p^{4}\rangle$), the initial value of the Patterson function of x-ray crystallography ($W_{3}$, $\langle p^{-3}\rangle$), the total electron-electron repulsion energy ($\langle p^3\rangle$), etc. Moreover, the position and momentum moments can be experimentally extracted as discussed  elsewhere \cite{gadre2,Thakkar1, Thakkar2, Lehtola}.

These ordinary and frequency moments play a relevant role  in the analysis of the structure and dynamics of natural systems and phenomena, from atomic and molecular systems to systems with non-standard dimensionalities, as can be seen in the excellent monographs of Dong \cite{dong}, Herschbach et al \cite{herschbach} and Sen \cite{sen}.

This work deals with some generalized position-momentum uncertainty relations which go far beyond the familiar uncertainty relation based on the standard deviation. By now, it is well known that the standard deviation is not at all the best measure of uncertainty because at times it cannot capture the essence of the uncertainty principle. The standard deviation is a reasonable measure of the spread of a probability distribution with a single hump (e.g., the gaussian and quasi-gaussian distributions). However, when the probability distribution has more than one hump, the standard deviation loses some of its usefulness, especially in connection with the notion of uncertainty. This problem is caused by the fact that the standard deviation attributes an ever increasing weight to the tails of the probability distribution; thus, a very slight contribution to the probability density, provided that it is located very far from the center, may cause the standard distribution to blow up. These observations have been reiteratively pointed out by various authors (see e.g. \cite{hilgevoord,maassen,majernik,uffink,bosyk}.

Accordingly, a variety of alternative formulations have been proposed which are based on other spreading measures of the probability distributions such as the ordinary moments of order higher orders and the frequency moments \cite{gadre1, Angulo1,angulo2,Zozor,Porras1,Porras2,maassen,sen}. Although endless variations on this theme can be given, let us just mention one practical application of these uncertainty inequalities: the problem of estimating the ground state energy for some given Hamiltonian. This technical problem has almost created an entire branch of mathematical physics, as can be seen in \cite{Lieb} and references therein. Needless to say, on the other hand, that lower and upper bounds for the products of moments in the two conjugate position and momentum spaces are very useful and relevant because, among many other things, they describe physical quantities which are experimentally accesible; in addition, the momentum-space quantities are not directly accessible, either in principle or due to experimental impediments.

Based on numerous semiclassical and Hartree-Fock-like ground-state calculations in atoms and diatomic molecules \cite{Pathak, Thakkar1, Hart-Thakkar, Thakkar-pedersen}, it has been found approximate relationships and semiclassical bounds connecting the momentum ordinary moments and position entropic moments of the form
\begin{equation}
\label{eq:empiricavevalmom2}
\langle p^{k}\rangle \leq c_{k} W_{1+\frac{k}{3}}[\rho] \quad \text{for} \quad k=-2,-1
\end{equation}
and
\begin{equation}
\label{eq:empiricavevalmom3}
\langle p^{k}\rangle \geq c_{k} W_{1+\frac{k}{3}}[\rho] \quad \text{for} \quad k=1,2,3,4
\end{equation}
with $c_{k}=3(3\pi^{2})^{k/3}(k+3)^{-1}$. Moreover the case $k = 2$ was already conjectured by Lieb, and weaker versions of it have been rigorously proved, as discussed elsewhere \cite{Thakkar-Pedersen}. These semiclassical bounds, which were found to be fulfilled by a large diversity of ground-state atoms and molecules \cite{Thakkar-Pedersen, Porras1, Porras2}, can be extended to $d$-dimensional systems of $N$ fermions with spin $s$ as
\begin{equation}
\label{eq:avevalmom4}
\langle p^{k}\rangle \geq K_{d}(k)q^{-\frac{k}{d}} W_{1+\frac{k}{d}}[\rho],
\end{equation}
where $k>0$, $q = 2s + 1$ and
\begin{equation}
\label{eq:Kdk}
 K_{d}(k) = \frac{d}{k+d}(2\pi)^{k}\frac{\left[\Gamma\left(1+\frac{d}{2}\right)\right]^{k/d}}{\pi^{k/2}}.
\end{equation}

And for $k<0$ the sign of inequality (\ref{eq:avevalmom4}) is inverted. Note that expression (\ref{eq:avevalmom4}) simplifies to (\ref{eq:empiricavevalmom2})-(\ref{eq:empiricavevalmom3}) for $d =3$ and $s=1/2$, since then $K_{d}(k) = 2^{\frac{k}{3}}c_{k}$. In fact, Eq. (\ref{eq:avevalmom4}) with constant $K'_{d}(k) = K_{d}(k) \times B(d,k)$ with $B(d,k) = \left\{\Gamma\left(\frac{d}{k}\right)\inf_{a>0}\left[ a^{-\frac{d}{k}}\left(\int_{a}^{\infty}du\, e^{-u}(u-a)u^{-1}\right)^{-1}\right] \right\}^{-\frac{k}{d}}$ has been rigorously proved by Daubechies \cite{Daubechies}. Table I collects some values of the constant $B(d,k)$ in terms of $d$ and $k$.
\begin{table}[H]
\begin{center}
\setlength{\tabcolsep}{7pt}
\renewcommand{\arraystretch}{1.5}
\begin{tabular}{|c|c|c|c|c|}
\hline
\multicolumn{5}{|c|}{$B(d,k)$}\\
\hline
\diagbox{$k$}{$d$} & $1$ & $2$ & $3$ & $4$\\
\hline
$1$  & $0.165728$ & $0.405724$ & $0.537513$ & $0.618094$\\
\hline
$2$  & $0.021331$ & $0.165728$ & $0.303977$ & $0.405724$\\
\hline
$3$  & $0.002056$ & $0.061935$ & $0.165728$ & $0.262190$\\
\hline
$4$  & $0.000158$ & $0.021331$ & $0.086812$ & $0.165728$\\
\hline
\end{tabular}
\end{center}
\caption{$B(d,k)$ for different values of $d$ and $k$.}
\label{table:table1}
\end{table}

As well, a number of authors have published some rigorous $d$-dimensional bounds of the same type \cite{Lieb, Hundertmark} with much less accuracy.\\[2mm]
On the other hand, similar expressions have been found which depend not on any global spreading measure (like the moments $W_{\alpha}[\rho]$) but on measures of the position probability with a property of locality (because they depend on the gradient of $\rho$), like the \textit{translationally or shift-invariant Fisher information} $I_d[\rho]$. Indeed, Zumbach \cite{Zumbach} has found that
\begin{equation}
\label{eq:zumbachd}
\langle p^{2}\rangle \leq \frac{1}{2}\left[1+C_{d}\left(\frac{N}{q} \right)^{2/3}\right]I_{d}[\rho],
\end{equation}
where the non-optimal constant $C_d$ is given by
\begin{equation}
\label{eq:Cd}
C_d=(4\pi)^{2}\frac{5d^2}{d+2}\left(\frac{2}{d+2}\right)^{2/d}
\end{equation}
for $1 \leq d \leq 5$, and $I_d[\rho]$ denotes the shift-invariant Fisher information of the electron probability density for $d$-dimensional $N$-fermion systems defined \cite{Frieden} as
\begin{equation}
\label{eq:FIshInf}
I_d[\rho] = \int_{\mathbb{R}_{d}} \frac{|\vec{\nabla}_{d}\sqrt{\rho(\vec{r})}|^{2}}{\rho(\vec{r})}\, d^{d}r = 4\int_{\mathbb{R}_{d}} \left(\vec{\nabla}_{d}\sqrt{\rho(\vec{r})}\right)^{2}\, d^{d}r,
\end{equation}
where $\vec{\nabla}_d$ denotes the $d$-dimensional gradient operator given by
\begin{equation}
\vec{\nabla}_{d}=\frac{\partial}{\partial r} \hat{r} +\frac{1}{r} \sum^{d-2}_{i=1} \frac{1}{\prod^{i-1}_{k=1} \sin \theta_k } \frac{\partial}{\partial \theta_i} \hat{\theta_i} + \frac{1}{r \prod^{d-2}_{i=1} \sin \theta_i } \frac{\partial}{\partial \varphi} \hat{\varphi} \nonumber,
\end{equation}
where the symbol $\hat{a}$ denotes the unit vector associated to the corresponding coordinate. Notice that  for $d=3$ the constant is $C_3= 9(4\pi)^{2}\left(\frac{2}{5}\right)^{2/3}$, and the Fisher information $I_3[\rho] = 4\int_{\mathbb{R}_{3}} (\nabla\sqrt{\rho})^{2}\, d^{3}r$ denotes the standard Fisher information of real $N$-fermion systems \cite{Frieden}.\\
The one-dimensional shift-invariant Fisher information is the translationally invariant version of the one-dimensional parametric Fisher information so much used to establish the ultimate bounds on sensitivity of measurements, which is a major goal of the parametric estimation theory. The latter quantity refers to the information about an unknown parameter  θ in the probability distribution estimated from observed outcomes. Let us assume that we want to estimate a parameter $\theta$ doing $n$ measures in an experiment. These data, $\vec{y} \equiv \left\lbrace y_i\right\rbrace_{i=1}^{n}$, obey $y_i=\theta+x_i$ where $\vec{x} \equiv \left\lbrace x_i \right\rbrace_{i=1}^{n}$ are added noise values. The noise $\vec{x}$ is assumed to be intrinsic to the parameter $\theta$ under measurement ($\theta$ has a definite but unknown value). This system is specified by a conditional probability law $p_{\theta}(\vec{y} | \theta)=p(y_1,y_2, \cdots,y_n | \theta)$ and $\hat{\theta}(\vec{y} | \theta)$ is, on average, a better estimate of $\theta$ as compared to any of the data observables, $\hat{\theta}(\vec{y}) = \theta$. In this case, we can define the parametric Fisher information as
\begin{equation}
I \equiv \int \left[ \frac{\partial  \ln p_\theta (\vec{y}) | \theta}{\partial \theta} \right]^2 p_\theta (\vec{y} | \theta) d\vec{y},
\end{equation}
which fulfils the known Cr\'amer-Rao inequality $\sigma^2 \times I \geq 1$, where $\sigma^2$ is the mean-square error given by
\begin{equation}\label{eqI_capI0:simga}
\sigma^2 = \int \left[\hat{\theta}(\vec{y})- \theta \right]^2 p_\theta (\vec{y}) d\vec{y}.
\end{equation}
Then, the parametric Fisher information measures the ability to estimate a parameter; that is, it gives the minimum error in estimating $\theta$ from the given probability density $p (\vec{y} \left| \theta \right.)$. In the particular case of $n=1$, $p_\theta (\vec{y} | \theta )=p (y | \theta)$ and the fluctuations $x$ are invariant to the size of $\theta$, $p_\theta (y | \theta)=p_x (y - \theta)$ with $x=y-\theta$ (i.e. shift invariance); one has
\begin{equation}\label{eqI_capI0:fisher_Def_d1}
I=\int \left[ \frac{\partial  \ln p (x)}{\partial x} \right]^2 p(x) dx = \int \frac{\left[p'(x)\right]^2}{p (x)}  dx,
\end{equation}
which is the one-dimensional {\it translationally-invariant Fisher information}. The extension to d dimensions is given by expression (\ref{eq:FIshInf}). This quantity is a measure of the gradient content of the density, so that it is very sensitive to the fluctuations of the density. Then, it quantifies the narrowness or localization of the density; so, it is a measure of the system disorder. See e.g., the monograph of Frieden \cite{Frieden} and references therein for further details.\\[2mm]

Nowadays the notion of translationally-invariant Fisher information is playing an increasing role in numerous fields \cite{Frieden}, in particular, for many-electron systems, partially because of its formal resemblance with kinetic \cite{sears, massen:pla01, luo:jpa02, Frieden} and Weisz\"acker \cite{Parr-Yang, romer} energies. The translationally-invariant Fisher information, contrary to the Shannon entropy, is a local measure of spreading of the density $\rho(\vec{r})$. The higher this quantity is, the more localized is the density, the smaller is the uncertainty and the higher is the accuracy in estimating the localization of the particle. However, it has an intrinsic connection with Shannon entropy via the de Bruijn inequality \cite{dembo, cover_91} as well as a simple connection with the precision (variance $V\left[ \rho \right]$) of the experiments by means of the celebrated Cr\'amer-Rao inequality \cite{dembo, cover_91}, $I\left[ \rho \right]\times V\left[ \rho \right] \geq d^2$.

The notion of Fisher information has been shown to be very fertile to identify, characterize and interpret numerous phenomena and processes in atomic and molecular physics such as e.g., correlation properties in atoms \cite{nagy}, the most distinctive nonlinear spectroscopic phenomena (avoided crossings) of atomic systems in strong external fields  \cite{gonza}, the periodicity and shell structure in the periodic table of chemical elements \cite{romera:mp02} and the transition state and the bond breaking/forming regions of some specific chemical reactions \cite{lopezrosa:jctc10}, as well as to systematically investigate the origin of the internal rotation barrier between the eclipsed and staggered conformers of ethane \cite{esquivel:2011} and the steric effect \cite{dehesa:2011}.

Recently, much effort is being devoted to build up a mathematical formulation of the quantum uncertainty principle based upon the Fisher-information measures evaluated on the conjugate position and momentum spaces. Nowadays it remains a strongly controversial problem \cite{hall,dehesa06,sanchez3,dehesa6,dehesa07,Sanchez2,plastino}. First, it was conjectured \cite{hall} in 2000 that the position-momentum Fisher information product had the lower bound $I_{1}(\rho) I_{1}(\gamma) \geq 4$ for one-dimensional quantum systems with the position and momentum densities $\rho(x) = |\Psi(x)|^2$ and  $\gamma(p) = |\Phi(p)|^2$, being $\Phi(p)$ the Fourier transform of $\Psi(x)$. Later in 2006 it was proved \cite{dehesa06} that this conjecture only holds for all real, even, one-dimensional wavefunctions $\Psi(x)$. Then, in 2011 this result was rigorously generalized \cite{Sanchez2} as $I_{d}(\rho) I_{d}(\gamma) \geq 4 d^2$ for the $d$-dimensional systems provided that either the position wavefunction $\Psi(\vec{r})$ or the corresponding momentum–space wavefunction $\Phi(\vec{p})$ is real \cite{Sanchez2}.

In addition, it has been found \cite{dehesa07} that the uncertainty product $I_{3}(\rho) I_{3}(\gamma)$ can be explicitly expressed in terms of the Heisenberg product $\langle r^2\rangle \langle p^2\rangle$ for any three-dimensional central potential; even more, it is fulfilled that $I_{3}(\rho) I_{3}(\gamma) \geq f(l,m)$, where $f(l,m)$ is a known simple function of the orbital and magnetic quantum numbers, given by $l$ and $m$, respectiveley. Furthermore, let us also mention that the product of position and momentum Fisher information has been proposed \cite{hall} as a measure of joint classicality of quantum states, what has been recently used for wave packet and quantum revivals \cite{Romera4}.\\
For completeness let us mention that a natural extension to the classical parametric Fisher information mentioned above, has been coined as (parametric) quantum Fisher information (see e.g., the monographs \cite{helstrom,holevo}) and successfully applied to quantum statistical inference and estimation theory in various directions (see e.g. \cite{braunstein,luo,watanave,shamsi,marzolino,benatti} and references therein).

In this work, we will use the $d$-dimensional Daubechies-Thakkar and  Zumbach expressions, given by (\ref{eq:avevalmom4}) and (\ref{eq:zumbachd}) respectively, to obtain novel (moment-based) Heisenberg-like and Fisher-information-based uncertainty-like relations for d-dimensional systems of $N$ fermions with spin $s$ in sections II and III, respectively. These relations extend and generalize previous general and specific uncertainty results of similar type. In addition, the accuracy of these results for a large variety of neutral and singly-ionized atoms and molecules is examined.

\section{Heisenberg-like uncertainty relations}

Let us here obtain lower bounds on the Heisenberg-like uncertainty products $\langle r^{\alpha}\rangle \langle p^{k}\rangle$, with $\alpha \geq 0$ and $-2 \leq k \leq 4$ for $d$-dimensional $N$-electron systems by taking into account both spatial and spin degrees of freedom. First we derive the bounds based on position and momentum expectation values with positive order, and then the corresponding ones involving momentum expectation values with a negative order. These results extend, generalize and/or improve similar results from various authors (see, e.g. \cite{gadre1,gadre2,tao,tian,Romera1,Romera2,Dehesa1,angulo3,Guerrero} and references therein).

\subsection{Uncertainty products $\langle r^{\alpha}\rangle \langle p^{k}\rangle$, with $\alpha \geq 0$ and $0 \leq k \leq 4$}

We begin with the semiclassical lower bound on the momentum expectation value $\langle p^{k}\rangle$ given by Eqs. (\ref{eq:avevalmom4})-(\ref{eq:Kdk}) in terms of the position entropy moments $W_{1+\frac{k}{d}}[\rho]$. Then, we apply the variational method of Lagrange's multipliers described in Refs. \cite{dehesa88,dehesa89} to bound the entropic moments $W_{q}[\rho]$. Indeed, let us minimize the quantity $\int [\rho(\vec{r})]^{q}\, d^{d}r$ subject to the constraints $\langle r^{0} \rangle \equiv \int \rho(\vec{r})\, d^{d}r = N$ and $\langle r^{\alpha} \rangle = \int r^{\alpha}\rho(\vec{r})\, d^{d}r$, $\alpha >0$, by taking variations of the form
\[
\delta\left\{\int [\rho(\vec{r})]^{q}\, d^{d}r - \lambda\int r^{\alpha}\rho(\vec{r})\, d^{d}r -\mu\int \rho(\vec{r})\, d^{d}r \right\}=0,
\]
where $\lambda$ and $\mu$ are Lagrange multipliers. One finds that the minimizer solution is given by the density
\[
f(r) = \left\{\begin{array}{cc}
				C(a^{\alpha}-r^{\alpha})^{1/(q-1)}, & r\leq a,\\
				0,									& r>a\\
				\end{array}\right.
\]
where the values of the factor $C$ and the parameter $a$ are determined so that the two previous constraints are fulfilled. In fact, following the lines indicated in Refs. \cite{galvez,dehesa88,dehesa89}, one can show that the quantity
\[
\int [f(r)]^{q}\, d^{d}r  =  F\, \, \langle r^{\alpha}\rangle^{-\frac{d}{\alpha}(q-1)}N^{\frac{d}{\alpha}(q-1)+q}
\]
is a lower bound of the wanted entropic moment $W_{q}[\rho]$, where F is a known analytic function of the parameters $q, \alpha$ and $d$. Then, with $q = 1 + \frac{k}{d}$ one finally obtains the rigorous inequality
\begin{equation}
\label{eq:entropmomlb}
W_{1+\frac{k}{d}}[\rho] \geq F(d,\alpha,k)\langle r^{\alpha}\rangle^{-\frac{k}{\alpha}}N^{1+k \left(\frac{1}{\alpha }+\frac{1}{d}\right)},
\end{equation}
where
\begin{eqnarray}
\label{eq:F}
F(d,\alpha,k) & = & \frac{\left(1+\frac{k}{d}\right)^{1+\frac{k}{d}}\alpha^{1+\frac{2k}{d}}}{\left[\Omega_{d}B\left(\frac{d}{\alpha},2+\frac{d}{k}\right)\right]^{\frac{k}{d}}}\nonumber \\
& \times & \left\{\frac{k^{k}}{\left[\left(1+\frac{k}{d}\right)\alpha+k\right]^{\left(1+\frac{k}{d}\right)\alpha+k}}\right\}^{\frac{1}{\alpha}},
\end{eqnarray}
where $\Omega_{d}=\frac{2\pi^{d/2}}{\Gamma(d/2)}$ is the volume of the unit hypersphere.

Then, from Eqs. (\ref{eq:avevalmom4}) and (\ref{eq:entropmomlb}) we obtain the generalized Heisenberg-like uncertainty relation given by
\begin{equation}
\label{eq:genheisuncprod}
\langle r^{\alpha}\rangle^{\frac{k}{\alpha}}\langle p^{k}\rangle \geq \mathcal{F}(d,\alpha,k)\,q^{-\frac{k}{d}}N^{1+k \left(\frac{1}{\alpha }+\frac{1}{d}\right)},
\end{equation}
where $\mathcal{F}(d,\alpha,k) = K_{d}(k)F(d,\alpha,k)$. From this general inequality of $N$-fermion systems with spatial dimensionality $d$ and spin dimensionality $q=2s+1$, we can make numerous observations. First, the case $k=2$ has been recently found \cite{Irene} by means of the Lieb-Thirring inequality. Second, there exists a delicate balance between the contributions of the spatial and spin degrees of freedom making the relation more or less accurate than the corresponding spinless inequality for either small or large d, respectively. Third, for $d=3$ and $q=2$ we obtain
\begin{equation}
\label{eq:genheisuncprod2}
\langle r^{\alpha}\rangle^{\frac{k}{\alpha}}\langle p^{k}\rangle \geq \mathcal{F}(3,\alpha,k)\,2^{-\frac{k}{3}}N^{\frac{k}{\alpha }+\frac{k+3}{3}},
\end{equation}
which holds for all N-electron systems. In particular, for $\alpha=k=2$ one has $\langle r^{2}\rangle\langle p^{2}\rangle \geq 1.85733 \times q^{-\frac{2}{3}}N^{\frac{8}{3}} = 1.17005 N^{\frac{8}{3}}$. A number of other Heisenberg-like relations, which are also instances of this inequality, is explicitly given in Table II.

\begin{table*}
\begin{tabular}{|c|c|c|c|c|}
\hline
\multicolumn{5}{|c|}{$$}\\
\multicolumn{5}{|c|}{$\langle r^{\alpha}\rangle^{\frac{k}{\alpha}}\langle p^{k}\rangle\geq f(N)$}\\
\multicolumn{5}{|c|}{$$}\\
\hline
\diagbox{$\alpha$}{$k$} &  $1$ & $2$ & $3$ & $4$\\
\hline
& & & &\\
\hspace{0.2cm}$1$ \hspace{0.2cm} &\hspace{0.1cm} $\frac{9}{49} (45\pi)^{1/3}N^{7/3}$\hspace{0.1cm} &\hspace{0.1cm} $\frac{243}{5324} (35\pi)^{2/3}N^{11/3}$\hspace{0.1cm} &\hspace{0.1cm} $\frac{243}{625}\pi N^{5}$\hspace{0.1cm} & \hspace{0.1cm}$\frac{841995}{39617584} (3465\pi^{4})^{1/3}N^{19/3}$\\
& & & &\\
\hline
& & & &\\
\hspace{0.2cm}$2$ \hspace{0.2cm} & \hspace{0.1cm} $\frac{9}{22}\sqrt{\frac{3}{11}}(35\pi)^{1/3}N^{11/6}$ \hspace{0.1cm}& \hspace{0.1cm} $\frac{9}{16}3^{2/3}N^{8/3}$\hspace{0.1cm} & \hspace{0.1cm} $\frac{135}{196}\sqrt{\frac{3}{7}}\pi N^{7/2}$ \hspace{0.1cm}& \hspace{0.1cm} $\frac{2268}{28561}\left(\frac{21}{13}\pi^{2}\right)^{1/3}\frac{\Gamma\left(\frac{17}{4}\right)}{\Gamma\left(\frac{11}{4}\right)}N^{13/3}$\\
& & & &\\
\hline
& & & &\\
\hspace{0.3cm}$3$ \hspace{0.3cm} &\hspace{0.1cm} $\frac{3}{5}\left(\frac{9}{5}\pi\right)^{1/3}N^{5/3}$\hspace{0.1cm} & \hspace{0.1cm} $3\left(\frac{45\pi}{196\sqrt{7}}\right)^{2/3}N^{7/3}$ \hspace{0.1cm}& \hspace{0.1cm} $\frac{1}{2}\pi N^{3}$\hspace{0.1cm} & \hspace{0.1cm} $\frac{189}{484}\left(\frac{63}{44}\pi^{4}\right)^{1/3}N^{11/3}$\\
& & & &\\
\hline
& & & &\\
\hspace{0.2cm}$4$\hspace{0.2cm} & \hspace{0.1cm} $\frac{3}{38}\left(\frac{3}{19}\right)^{1/4}(3465\pi)^{1/3}N^{19/12}$\hspace{0.1cm} & \hspace{0.1cm} $\frac{24\sqrt{3}}{169}\left(\frac{4\pi}{\sqrt{13}}\right)^{1/3}\left[\frac{\Gamma\left(\frac{17}{4}\right)}{\Gamma\left(\frac{3}{4}\right)}\right]^{2/3}N^{13/16}$ & $\frac{21}{4}\left(\frac{3}{11}\right)^{7/4}\pi N^{11/4}$\hspace{0.1cm} & \hspace{0.1cm} $\frac{567}{3200}\left(\frac{63}{2}\right)^{1/3}\frac{\pi^{2}}{\left[\Gamma\left(\frac{3}{4}\right)\Gamma\left(\frac{11}{4}\right)\right]^{4/3}} N^{10/3}$\\
& & & &\\
\hline
\end{tabular}
\caption{Some generalized Heisenberg-like uncertainty relations for $N$-electron systems, where both spatial and spin degrees of freedom are taking into account.}
\label{table:table2}
\end{table*}

\begin{figure*}
\minipage{0.32\textwidth}
\label{fig:neutros_rp}
\includegraphics[width=\linewidth]{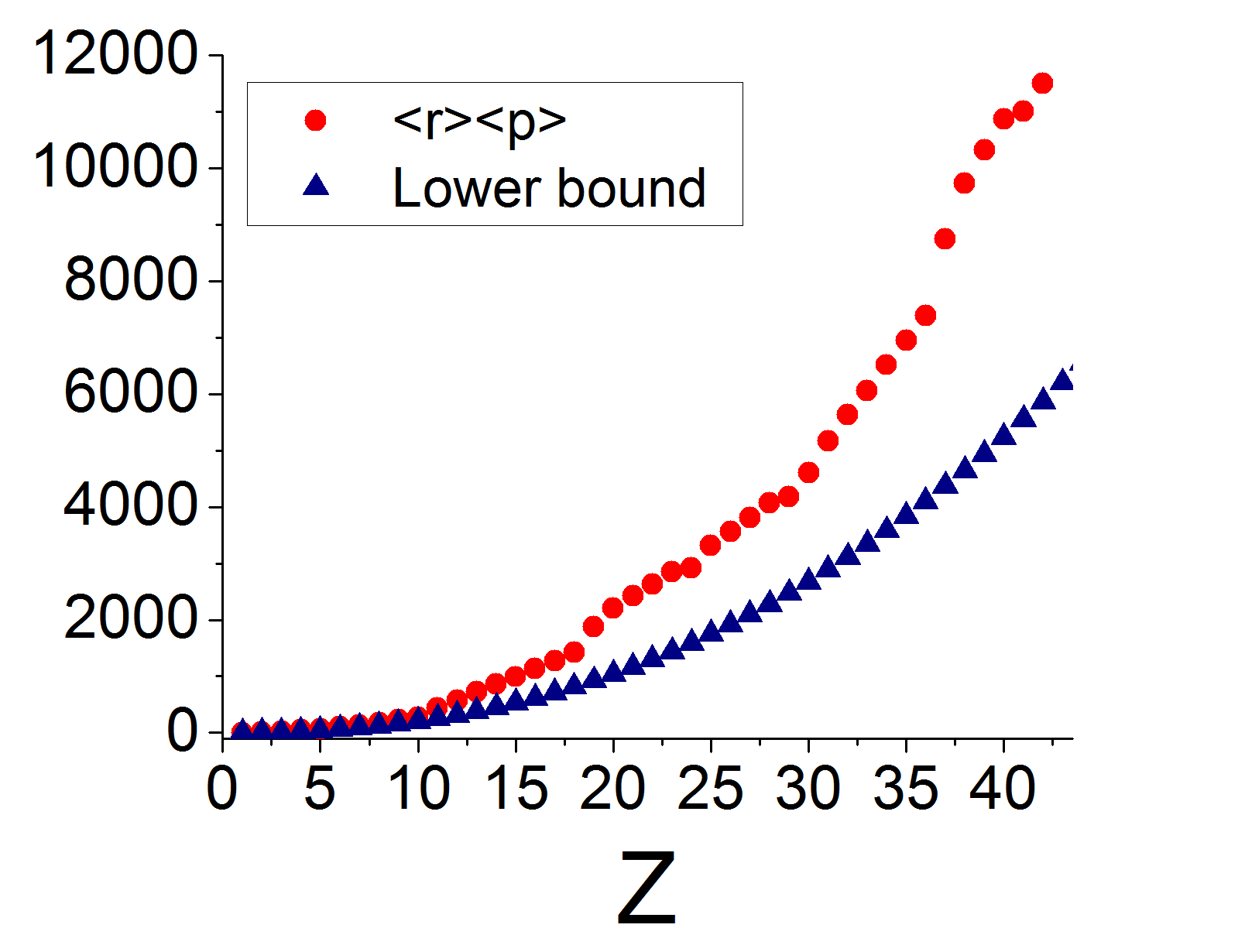}
\endminipage\hfill
\minipage{0.32\textwidth}
  \includegraphics[width=\linewidth]{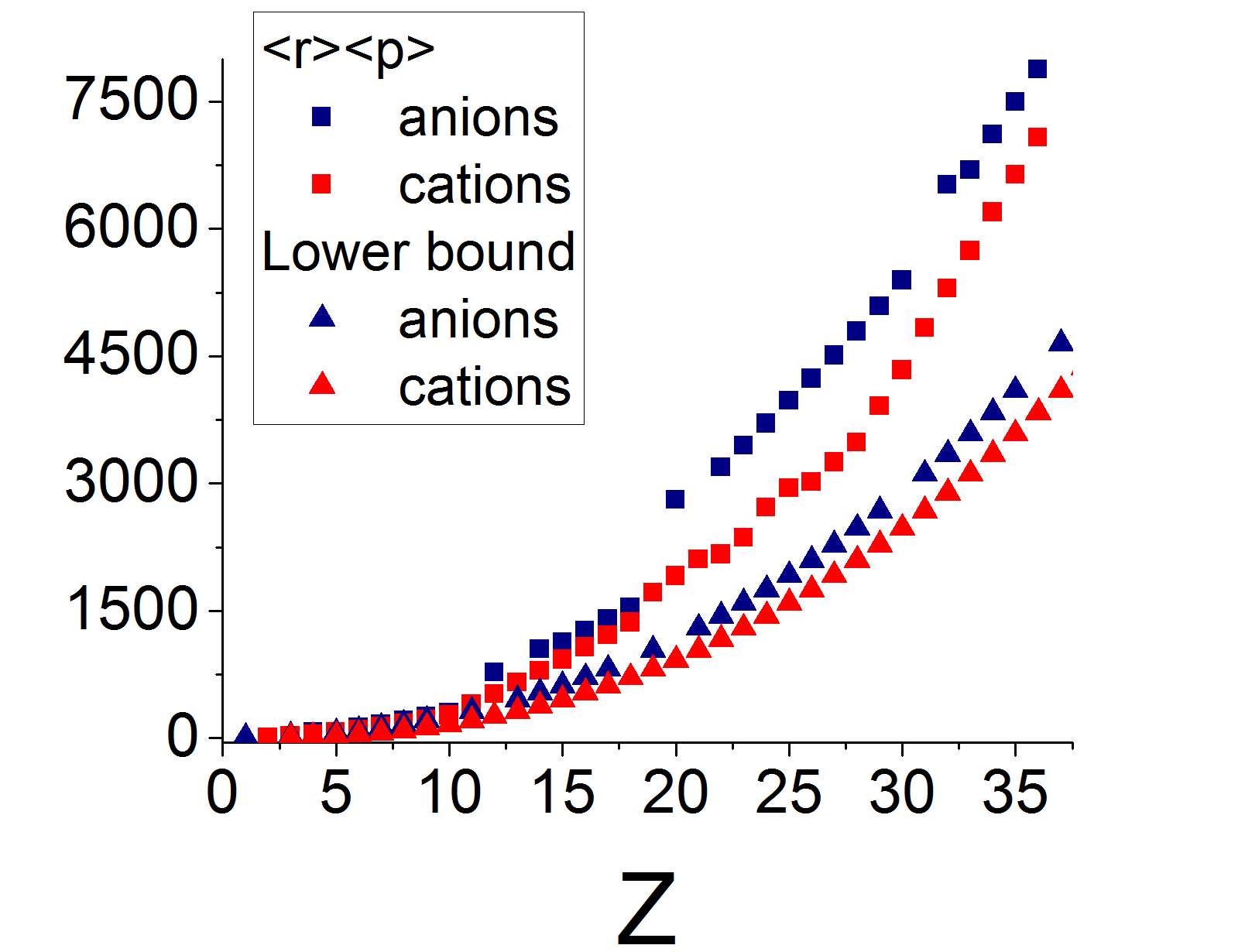}
  \label{fig:iones_rp}
\endminipage\hfill
\minipage{0.32\textwidth}
  \includegraphics[width=\linewidth]{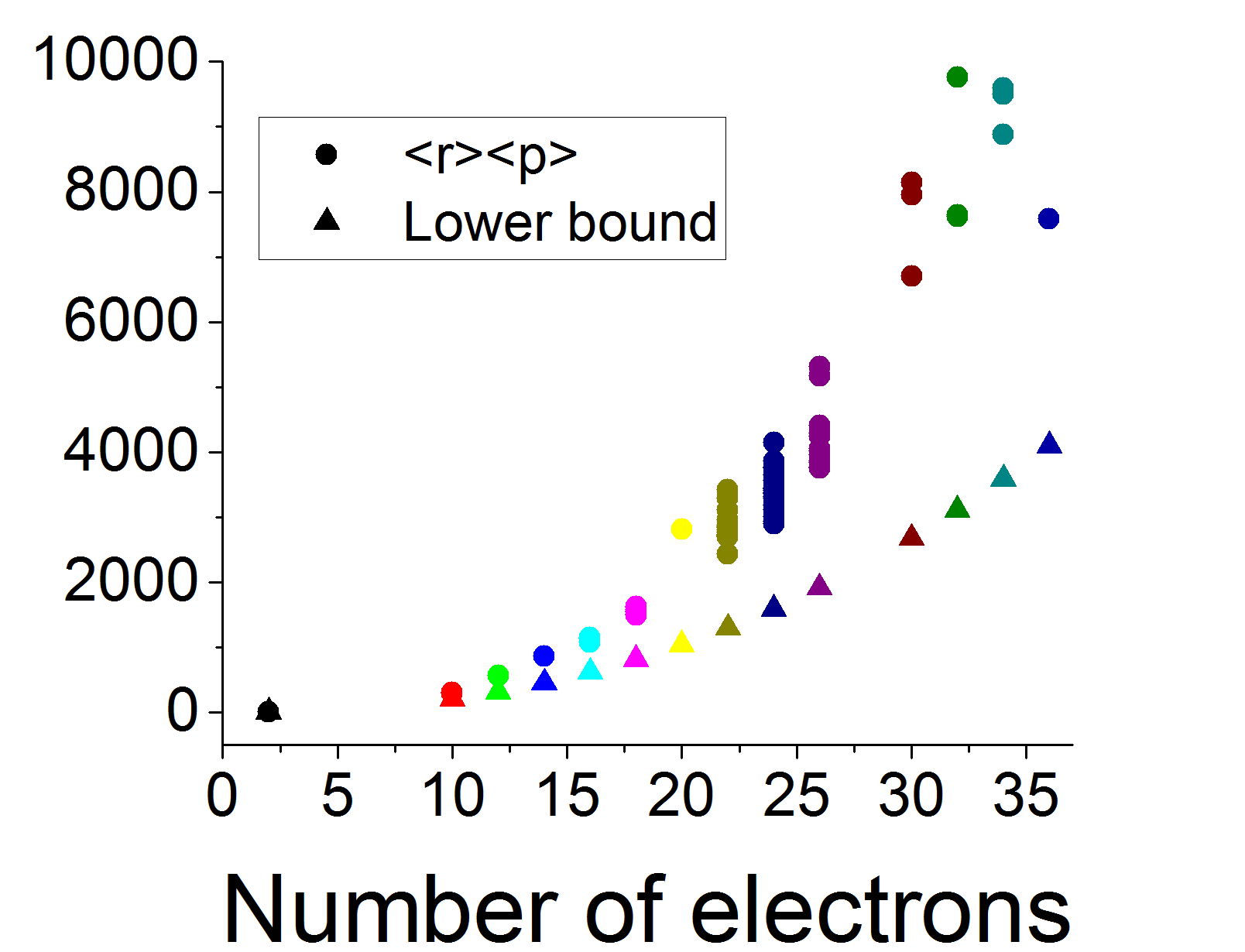}
  \label{fig:moleculas_rp}
\endminipage
\caption{(Color on line) Accuracy of $\langle r\rangle \langle p\rangle$ for all neutral atoms (left), all singly-ionized atoms (center) and $87$ polyatomic molecules (right). The symbol Z denotes the nuclear charge for atoms and ions. The colors in the molecular graph on the right correspond to different isoelectronic groups as explained in Appendix.}
\label{fig:1}
\end{figure*}

\begin{figure*}
\minipage{0.32\textwidth}
  \includegraphics[width=\linewidth]{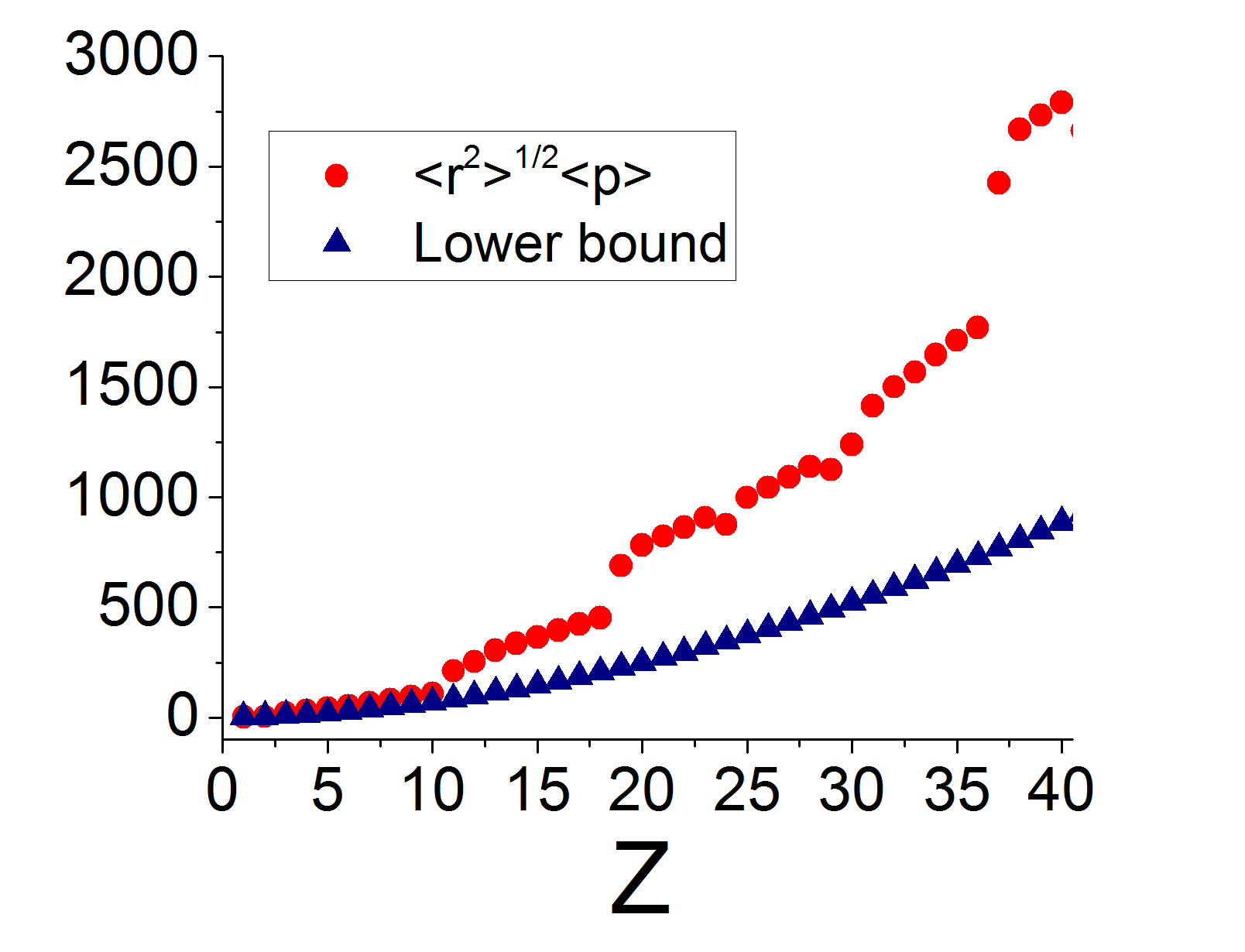}
  \label{fig:neutros_r2p}
\endminipage\hfill
\minipage{0.32\textwidth}
  \includegraphics[width=\linewidth]{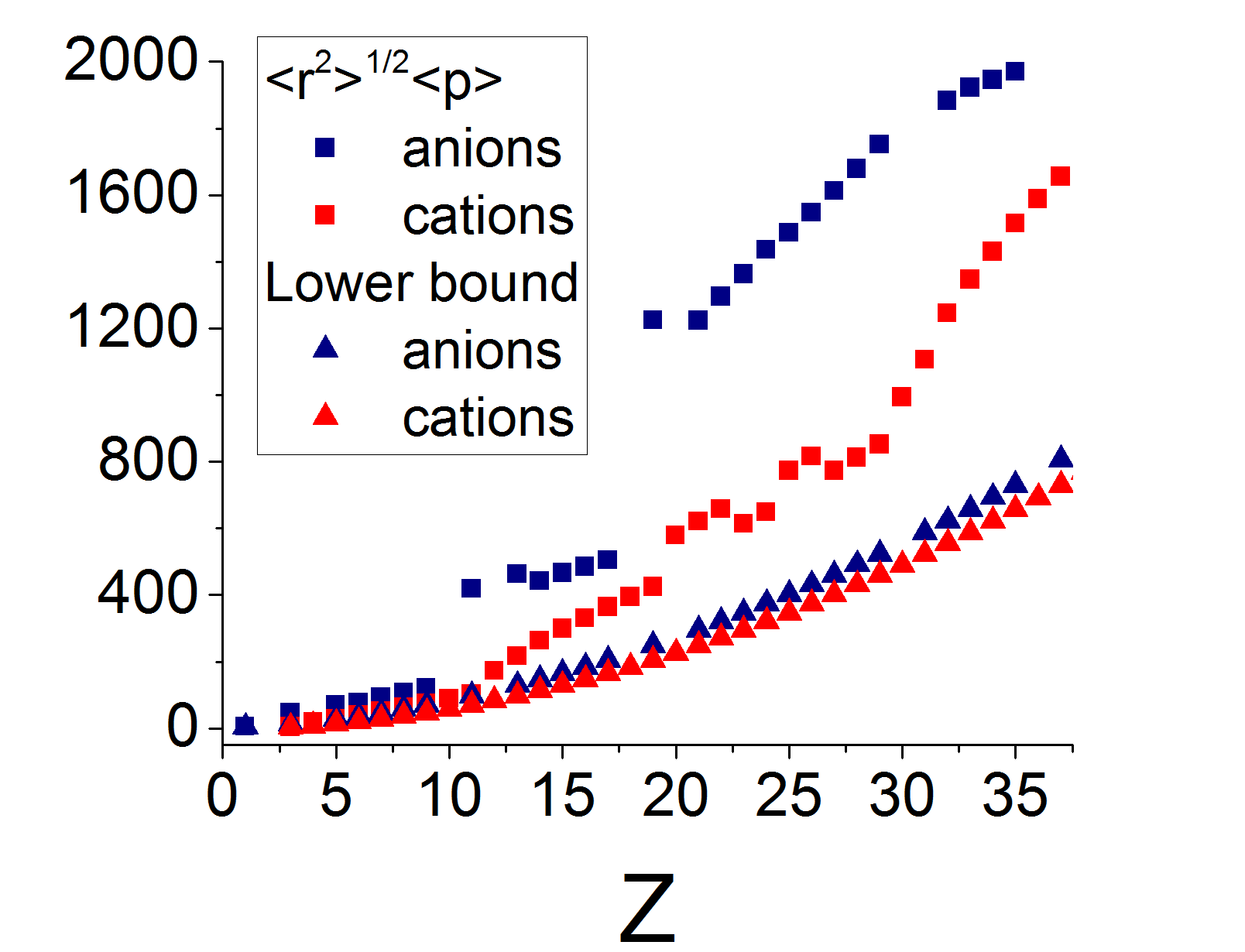}
  \label{fig:iones_r2p}
\endminipage
\minipage{0.32\textwidth}
  \includegraphics[width=\linewidth]{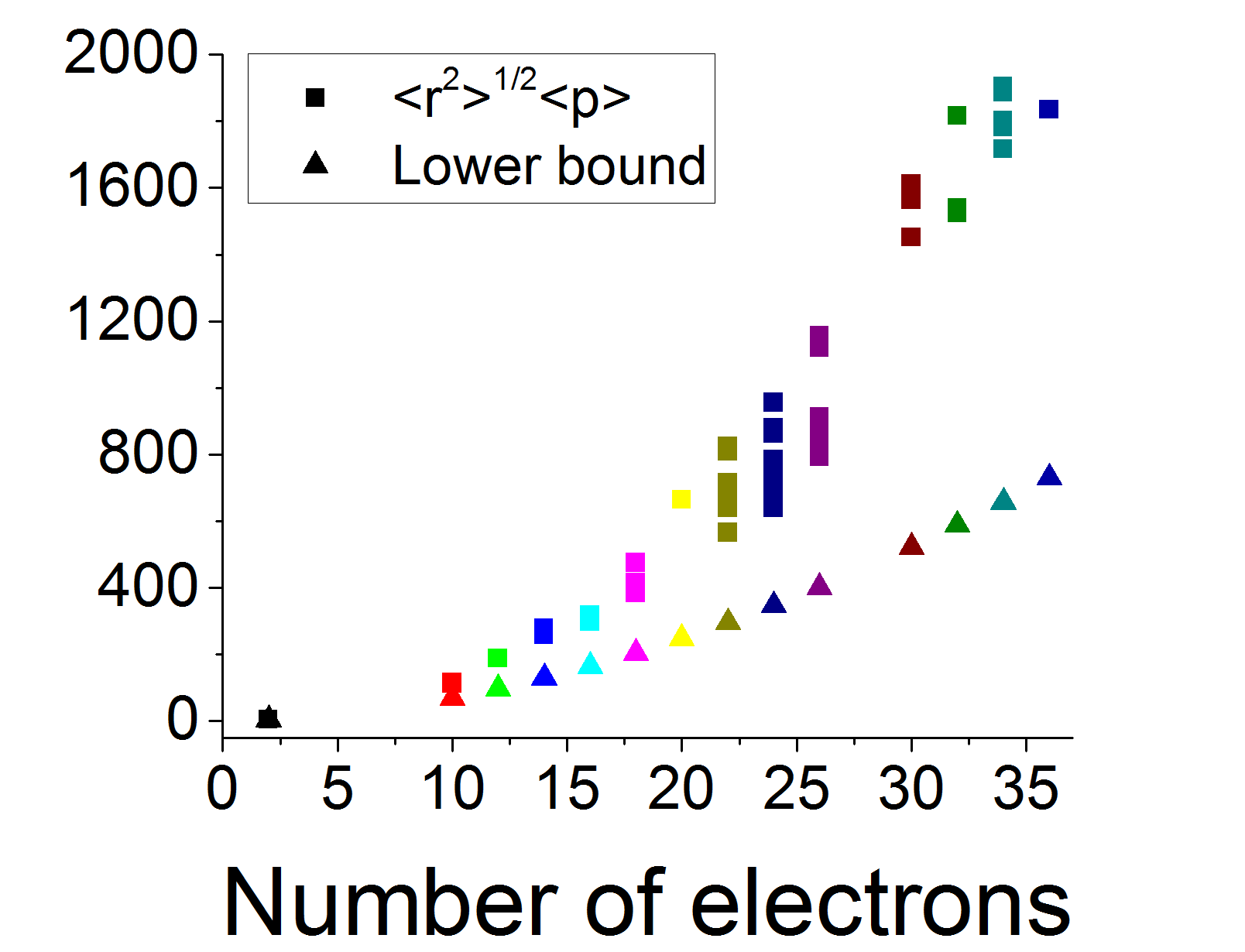}
  \label{fig:moleculas_r2p}
\endminipage\hfill
 \caption{(Color on line) Accuracy of $\langle r^{2}\rangle^{1/2} \langle p\rangle$ for all neutral atoms (left), all singly-ionized atoms (center) and $87$ polyatomic molecules (right). The symbol Z denotes the nuclear charge for atoms and ions. The colors in the molecular graph on the right correspond to different isoelectronic groups as explained in Appendix.}

 \label{fig:2}
\end{figure*}

Let us now study the accuracy of the uncertainty relation (\ref{eq:genheisuncprod}) for some values of $\alpha$ and $k$ in a large set of $N$-electron systems of neutral and singly-ionized atoms as well as in a variety of molecules. This is done in Fig.\ref{fig:1} and Fig.\ref{fig:2} for the Heisenberg-like products $\langle r\rangle \langle p \rangle $ and $\langle r^{2}\rangle^{1/2} \langle p \rangle $, respectively, for all ground-state neutral atoms of the periodic table from Hydrogen  ($N=1$) to Lawrencium ($N=103$) and their corresponding anions and cations, as well as for $87$ polyatomic molecules (see Appendix). The molecular set chosen for the numerical study includes different types of
chemical organic and inorganic systems (aliphatic and aromatic hydrocarbons, alcohols, ethers, ketones). It represents a variety of closed shell systems, radicals, isomers as well as molecules with heavy atoms such as sulphur, chlorine, magnesium and phosphorous. The symbol Z in both figures denotes the nuclear charge for atoms and ions. The colors in the molecular graph on the right of the two figures correspond to different isoelectronic groups described in Appendix.

The accurate near-Hartree-Fock wavefunctions of Koga et al \cite{koga1,koga2} have been used to evaluate the atomic uncertainty products.  In the molecular case we have used the Gaussian 03 suite of programs \cite{gaussian} at the $CISD/6-311++G(3df,2p)$ level of theory. For this set of molecules we have calculated position and momentum moments defined previously by employing software developed in our laboratory along with 3D numerical integration routines \cite{perezjorda} and the DGRID suite of programs \cite{kohout}.\\
For each figure the numerical values of these uncertainty products and the corresponding bounds (as given by Table I) are represented in terms of the number of electrons of the system under consideration. We first observe that the Heisenberg-like relations are indeed fulfilled in all cases, what is a check of our theoretical results. Then, we notice that our bounds are quite accurate for light electronic systems. Moreover, their accuracy decreases as the number of electrons increases. So, there is still a lot of space for improvement in heavy N-electron systems.\\

\subsection{Uncertainty products $\langle r^{\alpha}\rangle \langle p^{k}\rangle$, with $\alpha \geq 0$ and $k \leq 0$}

Here we start from the semiclassical lower bound on the momentum expectation value $\langle p^{k}\rangle$ given by Eqs. (\ref{eq:avevalmom4})-(\ref{eq:Kdk}) duly inverted because now k is assumed to have negative values, so that we have the following upper bound
\begin{equation}
\label{eq:uppboundp}
\langle p^{k}\rangle \leq K_{d}(k)q^{-\frac{k}{d}} W_{1+\frac{k}{d}}[\rho],
\end{equation}
in terms of the position entropy moments $W_{1+\frac{k}{d}}[\rho]$. Now, we use the above-mentioned variational method of Lagrange's multipliers given in Refs. \cite{galvez,dehesa88,dehesa89} to bound the entropic moments $W_{k'}[\rho]$ with the given constraints $\langle r^{0} \rangle = N$ and $\langle r^{\alpha} \rangle$, $\alpha < 0$, obtaining the rigorous inequality
\begin{equation}
\label{eq:uppbounentromom}
W_{k'}[\rho] \leq G_{d}(\alpha,k') \langle r^{\alpha} \rangle^{-\frac{k'}{\alpha}}N^{1+k' \left(\frac{1}{\alpha }+\frac{1}{d}\right)},
\end{equation}
where $k'<1$, $\alpha >\frac{d(1-k')}{k'}$, and
\begin{eqnarray}
\label{eq:Gfunc}
\hspace{-1cm}&&G_{d}(\alpha,k') =  \alpha^{1+\frac{2 k'}{d}} (-k')^{k'/\alpha} \left(\frac{1}{\alpha +\frac{\alpha\, k'}{d}+k'}\right)^{k' \left(\frac{1}{\alpha }+\frac{1}{d}\right)+1} \nonumber \\
&\times& \left(\frac{k'}{d}+1\right)^{\frac{k'}{d}+1} \left(\Omega_{d} \,B\left(-1-\frac{d (k'+\alpha )}{k' \alpha },\frac{d}{\alpha }\right)\right)^{-\frac{k'}{d}}
\end{eqnarray}
where again $\Omega_{d}=\frac{2\pi^{d/2}}{\Gamma(d/2)}$.

Finally, from Eqs. (\ref{eq:uppbounentromom}) and (\ref{eq:uppboundp}) we obtain in an algebraic manner the Heisenberg-like uncertainty relation
\begin{equation}
\label{eq:negpowuncerprod}
\langle r^{\alpha}\rangle^{\frac{k}{\alpha}}\langle p^{k}\rangle \leq \mathcal{G}_{d}(\alpha,k) q^{-k/d}N^{1+k \left(\frac{1}{\alpha }+\frac{1}{d}\right)},
\end{equation}
with  $k < 0$,  $\alpha > -\frac{3k}{k+d}$, and $\mathcal{G}_{d}(\alpha,k)=K_{d}(k)G_{d}(\alpha,k)$, for $d$-dimensional systems of $N$ fermions with spin $s$.

This fermionic inequality gives rise to the two following uncertainty relations
\begin{equation}
\label{eq:negpowuncerprod1}
\langle r^{\alpha}\rangle^{-\frac{1}{\alpha}}\langle p^{-1}\rangle \leq \mathcal{G}_{3}(\alpha,-1)\, 2^{1/3}N^{\frac{2}{3}-\frac{1}{\alpha}}, \quad \alpha >\frac{3}{2},
\end{equation}
and
\begin{equation}
\langle r^{\alpha}\rangle^{-\frac{2}{\alpha}}\langle p^{-2}\rangle \leq \mathcal{G}_{3}(\alpha,-2)\, 2^{1/3}N^{1+k \left(\frac{1}{\alpha }+\frac{1}{d}\right)}, \quad \alpha > 6,\nonumber
\end{equation}
for real $N$-electron systems, since then we have $d=3$ and $q=2$ and the exact $\langle p^{k}\rangle$ which are finite require that $k \geq -2$. As particular cases we have the Heisenberg-like uncertainty relations
\begin{eqnarray}
\label{eq:uncerprod2}
\langle r^{2}\rangle^{-\frac{1}{2}}\langle p^{-1}\rangle &\leq&  3^{\frac{1}{6}}2^{\frac{1}{3}}N^{\frac{1}{6}}\approx 1.51309 N^{\frac{1}{6}} ,\\
\langle r^{3}\rangle^{-\frac{1}{3}}\langle p^{-1}\rangle &\leq&  \left(\frac{6}{\pi}\right)^{\frac{1}{3}}N^{\frac{1}{3}}\approx 1.2407 N^{\frac{1}{3}} ,\\
\langle r^{4}\rangle^{-\frac{1}{4}}\langle p^{-1}\rangle &\leq& 2^{\frac{1}{2}} \left(\frac{3}{5}\right)^{\frac{5}{12}}N^{\frac{5}{12}}\approx 1.14308\,N^{\frac{5}{12}}
\end{eqnarray}
by making $\alpha = 2, 3$ and $4$, respectively in Eq. (\ref{eq:negpowuncerprod1}).

\section{Fisher-information-based uncertainty relation}

In this section we first express the position-momentum Fisher information product $I_{d}(\rho) I_{d}(\gamma)$ in terms of the Heisenberg uncertainty product $\langle r^2\rangle \langle p^2\rangle$ for $N$-electron $d$-dimensional systems. Then we use some results of the previous section to obtain a mathematical formulation of the position-mometum uncertainty principle for these systems. The resulting expressions extend and generalize various similar conjectures and inequalities in the sense already discussed in the first section \cite{hall,dehesa06,dehesa07,Sanchez2,plastino}.

We begin with Eq. (\ref{eq:zumbachd}) and, due to the reciprocity of the position and momentum spaces, its \textit{conjugate} inequality given by
\begin{equation}
\label{eq:r2ineq}
\langle r^{2}\rangle\leq  \frac{1}{2} \left[1+C_d\left(\frac{N}{q}\right)^{2/d}\right]I_d[\gamma],
\end{equation}
which lead to
\begin{equation}
\label{eq:fishprod1}
I_d[\rho]I_d[\gamma] \geq \frac{4}{\left[1+C_d\left(\frac{N}{q}\right)^{2/d}\right]^{2}}\langle r^{2}\rangle \langle p^{2}\rangle,
\end{equation}
This expression clearly manifests the uncertainty-like character of the product of the position Fisher information and momentum Fisher information for $N$-fermion systems. Moreover, let us now take into account  Eq. (\ref{eq:genheisuncprod}) with $\alpha = k = 2$, which gives the $d$-dimensional Heisenberg product \cite{Irene}
\begin{equation}
\label{eq:ddimheisprod}
\langle r^{2}\rangle \langle p^{2}\rangle \geq A(2,d) q^{-2/d}N^{2+2/d},
\end{equation}
with
\[
A(2,d) = \left\{\frac{d}{d+1}[\Gamma(d+1)]^{1/d}  \right\}^{2}.
\]
Then, the combination of Eqs. (\ref{eq:zumbachd}) and (\ref{eq:ddimheisprod}) leads to the following lower bound on the position-momentum Fisher-information product of $N$-fermion $d$-dimensional systems
\begin{equation}
\label{eq:ddimfishprod}
I_{d}[\rho] I_{d}[\gamma] \geq 4 A(2,d)\frac{N^{2/d +2}q^{-2/d}}{\left[1+C_{d}\left(\frac{N}{q}\right)^{2/d}\right]^{2}}.
\end{equation}
For electronic systems ($q=2$) this position-momentum uncertainty relation has the form
\begin{equation}
\label{eq:ddimfishprodelectr}
I_{d}[\rho] I_{d}[\gamma] \geq \frac{N^{\frac{2}{d}+2}\,2^{2-\frac{2}{d}} }{\left[1+N^{2/d}\,80 \pi ^2 d^2 (d+2)^{-\frac{d+2}{d}}\right]^2} A(2,d).
\end{equation}
Let us note here that for systems with a sufficiently large number of constituents N so that $1+C_{d}\left(\frac{N}{q}\right)^{2/d}\approx C_{d}\left(\frac{N}{q}\right)^{2/d} $ we obtain
\begin{equation}
\label{eq:ddimfishprod2}
I_{d}[\rho] I_{d}[\gamma] \geq N^{2-\frac{2}{d}} q^{\frac{2}{d}}\frac{(d+2)^{\frac{4}{d}+2}}{25 \pi ^4 4^{\frac{2}{d}+3}d^4} A(2,d)
\end{equation}
for fermionic systems, and
\begin{equation}
\label{eq:ddimfishprodelect}
I_{d}[\rho] I_{d}[\gamma] \geq N^{2-\frac{2}{d}}\frac{(d+2)^{\frac{4}{d}+2}}{25 \pi ^4 4^{\frac{1}{d}+3} d^4} A(2,d).
\end{equation}
for electronic systems.\\
And for real (i.e., $d=3$) $N$-electron systems we obtain from Eqs. (\ref{eq:ddimfishprodelectr}) and (\ref{eq:ddimfishprodelect}) the uncertainty relation

\begin{equation}
\label{eq:fishprod3}
I_{3}[\rho]I_{3}[\gamma] \geq \frac{N^{8/3}}{\left(N^{2/3}\frac{144 \pi ^2 }{5^{2/3}}+1\right)^2}\frac{3^{8/3} }{4},
\end{equation}
which for large $N$ reduces as
\begin{equation}
\label{eq:fishprod5}
I_{3}[\rho]I_{3}[\gamma] \geq N^{4/3}\frac{5}{3072\pi^{4}}\left(\frac{5}{3}\right)^{1/3},
\end{equation}
where $\frac{5}{3072\pi^{4}}\left(\frac{5}{3}\right)^{1/3} \approx 0.0000198107$.

\section{Conclusions}

The (variance-based) Heisenberg-Kennard relation is known to be a weak (and, at times, misleading) mathematical formulation of the quantum uncertainty relation \cite{hilgevoord,majernik}. Stronger uncertainty-like relations based either on moments of order other than 2 \cite{Angulo1,angulo3,Zozor} or on some information-theoretic quantities have been developed. Among the latter ones, the entropic uncertainty relations based on the Shannon entropy and on the R\'enyi entropy are well known \cite{bbm,bbi,zovi,zopo}. However the Fisher-information-based uncertainty-like relation still represents a controversial problem \cite{hall,dehesa06,sanchez3,dehesa6,dehesa07,Sanchez2,plastino} since its conjecture in 2000 for one-dimensional systems.\\
In this paper we have first found a set of (moment-based) Heisenberg-like uncertainty relations which extend and generalize the previous similar encountered expressions by starting from the Daubechies-Thakkar relations, which were semiempirically found by Thakkar for (three-dimensional) atoms and molecules and rigorously proved by Daubechies for $d$-dimensional quantum systems. Hereafter we have studied its accuracy for a large set of quantum systems: all the neutral and singly-ionized atoms of the periodic table and a large diversity of polyatomic molecules.  Later, we have shown the uncertainty character of the product of the position and momentum Fisher information of finite fermionic systems by expressing it in terms of the Heisenberg-Kennard position-momentum product by means of an inequality-type relationship. Moreover, we have found a lower bound on this product in terms of the number $N$ of its constituents. This result is not only relevant from a fundamental point of view, but also because of its physical implications on e.g., the determination of nonclassicality measures for quantum states as previously discussed. Finally, we should point out though that the latter bound can be certainly improved because the Zumbach constant $C_d$ is non optimal.

\acknowledgments
This work was partially supported by the Projects
FQM-7276, FQM-207 and FQM-239 of the Junta de Andaluc\'ia and the grant
FIS2011-24540 and FIS2014-54497P of the Ministerio de Econom\'ia y Competitividad (Spain).

\appendix
\section{Set of molecules used}

The molecular set chosen for the study includes different types of chemical organic and inorganic systems (aliphatic compounds, hydrocarbons, aromatic, alcohols, ethers, ketones). The set represents a variety of closed shell systems, radicals, isomers as well as molecules with heavy atoms such as sulphur, chlorine, magnesium and phosphorous. The geometries needed for the single point energy calculations above referred were obtained from experimental data from standard databases \cite{data}. The molecular set might be organized by isoelectronic groups as follows:\\
N-2: $H_2$ (hydrogen)\\
N-10: $NH_3$ (ammonia) , $CH_4$ (methane), $HF$ (fluoride hydride)\\
N-12: $LiOH$ (lithium hydroxide)\\
N-14: $HBO$ (boron hydride oxide), $Li_2O$ (dilithium oxide)\\
N-15: $HCO$ (formyl radical), $NO$ (nitric oxide)\\
N-16: $H_2CO$ (formaldehyde), $NHO$ (nitrosyl hydride), $O_2$ (oxygen)\\
N-17: $CH_3O$ (methoxy radical)\\
N-18: $CH_3NH_2$ (methyl amine), $CH_3OH$ (methyl alcohol), $H_2O_2$ (hydrogen peroxide), $NH_2OH$ (hydroxylamine)\\
N-20: $NaOH$ (sodium hydroxide)\\
N-21: $BO_2$ (boron dioxide), $C_3H_3$ (radical propargyl), $MgOH$ (magnesium hydroxide), $HCCO$ (ketenyl radical)\\
N-22: $C_3H_4$  (cyclopropene), $CH_2CCH_2$ (allene), $CH_3CCH$ (propyne), $CH_2NN$ (diazomethane), $CH_2CO$ (ketene), $CH_3CN$ (acetonitrile), $CH_3NC$ (methyl isocyanide), $CO_2$ (carbon dioxide), $FCN$ (cyanogen fluoride), $HBS$ (hydrogen boron sulfide), $HCCOH$ (ethynol), $HCNO$ (fulminic acid), $HN_3$ (hydrogen azide), $HNCO$ (isocyanic acid), $HOCN$ (cyanic  acid), $N_2O$ (nitrous oxide), $NH_2CN$ (cyanamide)\\
N-23: $NO_2$ (nitrogen dioxide), $NS$ (mononitrogen monosulfide), $PO$ (phosphorus monoxide),$C_3H_5$  (allyl radical), $CH_3CO$ (acetyl radical)\\
N-24: $C_2H_4O$ (ethylene oxide), $C_2H_5N$ (aziridine),  $C_3H_6$ (cyclopropane), $CF_2$ (difluoromethylene), $CH_2O_2$ (dioxirane), $CH_3CHO$ (acetaldehyde), $CHONH_2$ (formamide), $FNO$ (nitrosyl fluoride), $H_2CS$ (thioformaldehyde), $HCOOH$ (formic acid), $HNO_2$ (nitrous acid) $NHCHNH_2$ (aminomethanimine), $O_3$ (ozone), $SO$ (sulfur monoxide)\\
N-25: $CH_2CH_2CH_3$ (npropyl radical), $CH_3CHCH_3$ (isopropyl radical), $CH_3OO$ (methylperoxy radical), $FO_2$ (dioxygen monofluoride), $NF_2$ (difluoroamino radical), $CH_3CHOH$ (ethoxy radical),$CH_3S$ (thiomethoxy)\\
N-26: $C_3H_8$ (propane), $CH_3CH_2NH_2$ (ethylamine), $CH_3CH_2OH$ (ethanol), $CH_3NHCH_3$ (dimethylamine), $CH_3OCH_3$ (dimethyl ether), $CH_3OOH$ (methyl peroxide), $F_2O$ (difluorine monoxide)\\
N-30: $ClCN$ (chlorocyanogen), $OCS$ (carbonyl sulfide), $SiO_2$ (silicon dioxide)\\
N-31: $PO_2$ (phosphorus dioxide), $PS$ (phosphorus sulfide)\\
N-32: $ClNO$ (nitrosyl chloride), $S_2$ (sulfur diatomic), $SO_2$ (sulfur dioxide)\\
N-33: $ClO_2$ (chlorine dioxide), $OClO$ (chlorine dioxide)\\
N-34: $CH_3CH_2SH$ (ethanethiol), $CH_3SCH_3$ (dimethyl sulfide),$H_2S_2$ (hydrogen sulfide), $SF_2$ (sulfur difluoride)\\
N-36: $HBr$ (bromide hydride)\\

\end{document}